*In-situ* Thermal Transport Measurement of Flowing Fluid using Modulated Photothermal Radiometry

Jian Zeng[1*], Ka Man Chung[2*], Sarath Reddy Adapa[1], Tianshi Feng[1], Renkun Chen[#1,2]

[1]Department of Mechanical and Aerospace Engineering, University of California, San Diego, La Jolla, California 92093, United States

[2]Program of Material Science and Engineering, University of California, San Diego, La Jolla, California 92093, United States

[*]These authors contributed equally.

[#]Corresponding author: rkchen@ucsd.edu

**Abstract**

*In situ* thermal transport measurement of flowing fluid could be useful for the characterization and diagnosis of practical thermal systems such as fluid heat exchangers and thermal energy storage systems. Despite abundant reports on the *ex-situ* thermal conductivity measurement of stagnant fluids, a suitable technique for the thermal conductivity measurement of flowing fluid has been rarely reported. This paper presents the thermal conductivity measurement of flowing fluid within a pipe using a non-contact modulated photothermal radiometry (MPR) technique, where the surface of the pipe is heated by an intensity-modulated laser and the heat diffuses into the fluid with suitable modulation frequency. We design a tube section with small wall thickness suitable for the MPR measurements to maximize the sensitivity of the thermal response to the fluid properties while minimizing the lateral heat spreading effect. Intrinsic thermal conductivity of different fluids was obtained within a proper range of frequency and flow velocity where the forced convection effect is negligible. The forced convection effect became prominent at high flowing velocity and at low modulation frequency, leading to overestimated thermal conductivity of fluid. It is found that the intrinsic thermal conductivity could be obtained when the flow velocity is less than 100 mm/sec and $Re_D^{1/2}Pr^{1/3} < 100$ for DI water and Xceltherm oil under the specified experimental conditions, where $Re_D$ is the Reynolds number and $Pr$ is the Prandtl number.

**Keywords:** Photothermal Radiometry; Thermal conductivity; Flowing fluid; *In-situ* Measurement; Convection heat transfer.

## 1. Introduction

Measurement of thermal conductivity of fluids is important for the design of heat exchangers [1, 2], nuclear reactors [3, 4], thermal energy storage systems [5], and many other thermal systems. The ability of *in-situ* thermal conductivity measurement of flowing fluids could be useful for the operational safety and diagnosis of these systems [6, 7]. Despite a multitude of techniques available for the thermal conductivity measurement of stagnant fluids, including the transient hot-wire (THW) method [8, 9], steady-state method [10-12], laser flash analysis (LFA) [13, 14], 3ω technique[6, 15, 16], and time or frequency domain thermo-reflectance techniques (TDTR/FDTR)[17], there are still challenges of applying these techniques for *in-situ* diagnostics, especially under the harsh environment (such as high temperature and/or with corrosive fluids). Hong *et al.* applied the 3ω technique for the thermal conductivity measurement of flowing water and ethanol where a thin metallic transducer layer in contact with the fluid serves as both the heater and the thermometer. They obtained the thermal conductivity of fluids at low flow velocity [15]. The 3ω technique was also used for the measurement of thermal properties of flowing gases [12, 18]. Based on the similar principle, thermal conductivity of flowing fluids has been measured using the TDTR method[19]. Despite the successful implementation of the *in-situ* 3ω and TDTR measurements on common fluids, where dedicated metallic transducers are needed, it is still challenging to extend them for *in-situ* measurements under challenging conditions, such as at high temperature and for corrosive fluids (e.g., molten salts for nuclear reactors [20] and concentrated solar power (CSP) [21]). For example, to avoid the corrosion by molten salts to the metallic transducer in the 3ω method, a thin protective coating (e.g., alumina) has to be applied, which not only requires time-consuming microfabrication, but also poses question on the long-term durability[22]. Therefore, a suitable technique for the *in-situ* thermal conductivity measurement of flowing fluids under harsh environment would be desirable.

Photothermal radiometry (PTR) using either modulated (continuous wave) or pulsed laser is a well-known non-contact thermal characterization tool for bulk and coating materials [23-25]. In the modulated photothermal radiometry (MPR), a sample is heated by an intensity-modulated laser to probe into different depths of the sample and its surface temperature response is measured with an infrared (IR) detector [24, 26]. We have previously established the MPR technique for high-temperature measurements of bulk and thin coating samples, by using a refractory black coating for laser absorption and IR surface thermometry[25]. In this work, we applied the MPR technique for *in-situ* thermal conductivity measurement of flowing fluids within a pipe. By modulating the frequency of the laser and controlling the thermal penetration depth ($L_p = \sqrt{\frac{2\alpha}{\omega}}$ where $\alpha$ is thermal diffusivity of fluid and $\omega$ is the angular frequency) beyond the wall thickness of the tube and within the momentum/thermal boundary layer of the fluid, we could get high sensitivity of the surface temperature response to the thermal conductivity of the fluid while avoiding the convection effect within a certain range of flow velocity, thus obtaining the intrinsic thermal conductivity of the flowing fluid. We demonstrate the MPR measurement of flowing liquids (DI water, ethanol, Dowtherm A oil and Xcertherm 600 oil) in a pipe with flow velocity up to 550 mm s$^{-1}$. The Reynolds number ($Re_D$) is ranging from 0 to 13,000 at temperature from 30°C to 170°C, thus covering the stationary, laminar flow and turbulent flow regimes. Here $Re_D$ is defined based on the pipe diameter $D$. At low flow velocity and low $Re_D$, the intrinsic thermal conductivity of fluid is obtained. At higher velocity, the forced convection effect becomes more prominent, and the obtained effective thermal conductivity is higher. This transition is found to relate to the dimensionless number $Re_D^{1/2} Pr^{1/3}$, where $Pr$ is the Prandtl number. When $Re_D^{1/2} Pr^{1/3}$ is less than 100, we can obtain the intrinsic thermal conductivity of the flowing fluids. The MPR technique reported here can provide a facile *in-situ* thermal conductivity diagnostic tool for flowing fluids with broad applications. It may open an opportunity to simultaneously quantify both the thermal conductivity and local convective heat transfer coefficient in a fluid loop.

## 2. Experimental Mothed and Modelling

### 2.1. Experimental Section

**Figure 1a** displays the schematic of the MPR system integrated with a flowing liquid loop. The MPR system is the same as what was reported in our previous study [25] except that the test section is replaced with a circular tube connected to the fluid loop. A waveform-generator drove a continuous wave (CW) diode laser with its intensity modulated as sinusoidal function at the angular frequency of $\omega$. The laser spot size was controlled to be around 10 mm in diameter and the laser beam was homogenized from Gaussian beam into a top-hat profile. The test section was heated up by the modulated laser flux, leading to the oscillation of the surface temperature rise $\theta_s$ of the sample at the same frequency ($\omega$) as the heating laser. $\theta_s$ was measured based on the thermal emission from the surface collected by a HgCdTe (MCT) detector after proper calibration using a pyrometer. More details of the MPR system can be found in our previous work [25].

The schematic and photograph of the test section of the fluid flowing pipe are shown in **Figures 1b and 1c**, respectively. The wall thickness of the test section was set to be 100 μm to have high measurement sensitivity and accuracy as discussed in the later section. To ensure the mechanical strength of the setup, a 100 μm thick stainless steel 316 (SS316) sheet was wrapped into a round shape and welded onto two short tubes of 25.4 mm in outer diameter and 1 mm in wall thickness on both ends. The middle thin-walled section was ~10 cm long to minimize the edge effect due to the change in the wall thickness in this section. The laser spot was about 10 mm in diameter and was located at the center of the thin-walled region. The exterior surface of the tube was coated with a 30-μm-thick Pyromark 2500 black coating (see Figure S1 in Supplementary Information). During the MPR measurement, the laser beam irradiated the center of the middle thin-walled section. The tube was connected to the fluid loop with the fluid supplied by a pump from a liquid reservoir with heating capability (ZNCL Vevor 15843221637152051). Four types of fluids were measured: DI water, ethanol, Xceltherm 600 oil (Radco Industries Inc.), and Dowtherm A oil (Sigma Aldrich). Liquid was circulating at different flow velocities up to 550 mm s$^{-1}$ driven by a mechanical pump (HD Portable oil transfer pump, Massive gears 3 phase, Goldenstream Pumps) with a VFD module (ABB ACS 150 Drive) to control the flow rate, which was also monitored using a flow rate meter. The pump has an operational temperature limit of ~200 ºC, so our measurements were limited to 170 ºC.

## 2.2. Measurement Principle

**Figure 2a** shows the schematic of the MPR measurement of a flowing fluid inside a tube. The fluid is supplied from one end of the steel tube at a fixed bulk velocity $u_b$. An intensity-modulated laser at angular frequency of $\omega$ irradiates the surface of the middle thin-walled section. The surface temperature $\theta_s$ is oscillating at $\omega$. The thermal penetration depth ($L_p$) can be adjusted by changing $\omega$ of the incident laser. The thermal emission from the tube surface is collected by an IR detector to measure the $\theta_s$ after calibration with a pyrometer as described in detail in our earlier work [25]. **Figure 2b** shows the schematics of the momentum boundary layer $\Delta$, thermal boundary layer $\delta$ and the thermal penetration depth $L_p$ during the MPR measurement. Assuming laminar flow, the hydrodynamic entrance length $L_{ef}$ and the thermal entrance length $L_{eh}$ can be estimated by:

$$L_{ef} = 0.05 Re_D D, \tag{1}$$

and

$$L_{eh} = 0.033 Re_D Pr D, \tag{2}$$

where $D$ is the diameter of the tube. In our measurement, $Re_D$ is in the range of 100-13000 and $Pr$ is in the range of 5-300. Therefore, $L_{ef}$ and $L_{eh}$ are in the range of 13-1650 cm and 42-5450 cm, respectively. The laser spot (~10 mm diameter) of the MPR measurement is located at middle of the 10-cm long thin-walled section. Therefore, the fluid flow is still in the hydrodynamic and thermal entrance region at the position of the MPR measurement. In this region, the thermal boundary layer ($\delta$) is always thinner than the momentum boundary layer ($\Delta$) as they are related by $\delta = Pr^{-1/3} \Delta$ and $Pr^{-1/3}$ is 0.15-0.58 for the fluids to be measured. To measure the intrinsic thermal conductivity of a flowing fluid, the thermal penetration depth should be smaller than both the momentum boundary and thermal boundary layer thicknesses: $L_p <$ min $[\Delta, \delta]$. In our case, since $\delta$ is always smaller than $\Delta$, $L_p < \delta$ should be satisfied.

As shown in **Figure 2b**, at high frequency, $L_p$ is shorter than $\delta$ and thus the convection effect is minimized. In this case, the intrinsic thermal conductivity of the fluid can be obtained. On the contrary, at low frequency with $L_p > \delta$, the convection effect becomes prominent, leading to overestimation of the measured thermal conductivity of fluid. It is noted that $\delta$ depends on the flowing velocity and both the $\delta$ and $L_p$ depend on the thermophysical properties of the fluid and the temperature, so the desirable frequency range to obtain the intrinsic thermal conductivity may change with the fluid properties, velocity, and temperature. In addition, $L_p$ must be larger than the wall thickness of the tube (100 µm) such that the measurement is more sensitive to the fluid than the wall. Therefore, the frequency and $L_p$ must fall within a suitable range. In general, $L_p$ in the fluid is controlled to be around ~150 - ~300 µm in our experiments.

Although the measurement is conducted in a cylindrical tube, the 2-D heat transfer model based on the planar geometry can be applied (as justified later). Within the fluid, in the absence of heat generation or viscous dissipation and assuming constant thermal properties, the 2D heat transfer equation in the frequency domain for the flow in the entrance region is:

$$\frac{\partial^2 \theta_f}{\partial x^2} + \frac{\partial^2 \theta_f}{\partial y^2} = \frac{j\omega}{\alpha_f} \theta_f + \frac{\tilde{v}}{\alpha_f} \frac{\partial \theta_f}{\partial x}, \tag{3}$$

where $x$ is the coordinate in the direction along the fluid flow and $y$ is the direction perpendicular to the fluid flow as shown in **Figure 2b**. The position $y = 0$ is defined at the wall. $\theta_f$ is the temperature of the fluid, $\alpha_f$ is the thermal diffusivity of the fluid, $j$ is the imaginary unit, $\tilde{v}$ is the velocity vector of flowing fluid and $\omega$ is the angular frequency. The velocity profile within the laminar boundary layer assuming a cubic polynomial is:

$$\tilde{v} = u_b \left( \frac{3}{2} \frac{y}{\Delta} - \frac{1}{2} \left( \frac{y}{\Delta} \right)^3 \right), \tag{4}$$

where $u_b$ is the bulk velocity. Within a short distance from the wall, the flow pattern can be assumed to be linear, which is valid when $L_p \ll \Delta$. As such, equation 4 is reduced to:

$$\tilde{v} = \frac{3u_b}{2\Delta} y . \tag{5}$$

Therefore, equation 3 becomes:

$$\frac{\partial^2 \theta_f}{\partial x^2} + \frac{\partial^2 \theta_f}{\partial y^2} = \frac{j\omega}{\alpha_f} \theta_f + \frac{3u_b y}{2\alpha_f \Delta} \frac{\partial \theta_f}{\partial x} . \tag{6}$$

Clearly, for $y \to 0$ (near the wall), **equation 6** is reduced to a pure heat conduction equation without the advection term. Therefore, the convection effect is only important when the temperature field reaches deep into the boundary layer, i.e., at the low frequency where $L_p$ is large. By doing a dimensional analysis, it was found that the forced convection effect is negligible when the following condition is satisfied [15]:

$$\frac{u_b}{\omega} < L_c, \tag{7}$$

where $L_c$ is the characteristic length scale of the fluid flow. For a fully developed flow, such as the one studied in Ref. [15], $L_c$ is the radius of the flow channel. In our case studied here, the flow is still in the entrance region, so $L_c$ should be the momentum boundary layer thickness $\Delta$. If we take $L_c = \Delta$ and knowing $\Delta = 5\sqrt{\frac{vx}{u_b}}$ (for a laminar flow), **equation 7** becomes:

$$u_b < \left(5\sqrt{vx}\omega\right)^{\frac{2}{3}} \tag{8}$$

where $v$ is the kinematic viscosity of the fluid and $x$ is the characteristic entrance length of the measurement section as shown in Figure 1b ($x = 10$ cm in our experiment). **Figure 3a** shows the critical velocity as a function of $\omega$. The angular frequency is in the range of 3 – 12 rad s$^{-1}$ (i.e., $f = 0.5 - 2$ Hz and $L_p = 150 - 300$ µm), and thus the critical velocity is ~ 70 mm s$^{-1}$ to have the negligible forced convection effect. The critical velocity increases with increasing frequency because the penetration depth decreases with increasing frequency, and thus the effect of the momentum boundary layer diminishes.

The above analysis only examines the velocity and momentum equation. From the heat transfer point of view, $L_p < \delta$ also needs to be satisfied in order to measure the intrinsic thermal conductivity of the fluid, as discussed earlier. Here we set $L_p < 0.2\delta$:

$$\frac{L_p}{\delta} = \frac{\sqrt{\frac{\alpha_f}{\pi f}}}{5\sqrt{\frac{\nu x_1}{u_b}} Pr^{-1/3}} < 0.2, \tag{9}$$

where $x_1$ is the length of heater and is equal to the laser spot size, i.e., $x_1 = D_{laser}$. Rearranging **equation 9** leads to:

$$Re_D^{1/2} Pr^{1/3} < \sqrt{\frac{D D_{laser} \pi f}{\alpha_f}}, \tag{10}$$

where $D$ is the pipe diameter (25 mm), $Re_D = \frac{u_b D}{\nu}$, and $D_{laser} = 10\ mm$ in our experiment. Evaluating **equation 10** at $f = 1$ Hz, we have $Re_D^{1/2} Pr^{1/3} < \sim 100$ as the condition to neglect the forced convection effect. On the other hand, when $Re_D^{1/2} Pr^{1/3} > \sim 100$, the forced convection effect becomes important, and the measured effective thermal conductivity is expected to be higher than the intrinsic thermal conductivity of the fluid. In our measurements, we vary the flow velocity $u_b$, laser modulation frequency $f$, and temperature (thus the viscosity) of the fluid to access both the intrinsic thermal conductivity (pure heat conduction) and forced convection regimes from the fluid flow (**equation 8** and **figure 3a**) and the heat transfer (**equation 10**) points of view.

In addition, although the MPR measurement is conducted on a cylindrical tube, the heat transfer can be modeled based on the planar geometry under certain conditions. To specify these conditions, the thermal response of a cylindrical tube is compared to that of a flat plate. The surface temperature rise given by the heat transfer model for a cylindrical tube under the modulated laser heating is [27]:

$$\theta_{s,cyl}(r,\phi,\omega) = \frac{q_s}{2\pi k_s} \left( \begin{array}{l} \frac{2I_0(\sigma r)}{I_0'(\sigma D/2)} \sin\left(\frac{\theta_0}{2}\right) + \frac{I_1(\sigma r)}{I_1'(\sigma D/2)} (\theta_0 + \sin(\theta_0)) \times \cos\left(\frac{\pi}{2} - \phi\right) \\ + 2\sum_{m=2}^{\infty} \frac{I_m(\sigma r)}{I_m'(\sigma D/2)} \times \cos\left[\frac{m}{2}(\pi - 2\phi)\right] \times \left[ \frac{\sin\left[\frac{(m+1)\theta_0}{2}\right]}{m+1} + \frac{\sin\left[\frac{(m+1)\theta_0}{2}\right]}{m-1} \right] \end{array} \right) \quad (11)$$

where $k_s$ is the thermal conductivity of the solid tube, $\theta_0$ is the center angle of the laser spot and is $\pi/7.8$ for a laser spot diameter ($D_{laser}$ = 10 mm); $\phi$ is the angle of the detection spot ($D_{IR} = 1\ mm$) relative to the horizontal position and is $\pi/2$ if the laser and IR-detector are well aligned. $I_m$ is the $m^{th}$ order modified Bessel function of the first kind and $I_m'$ is the derivative of $I_m$. $\sigma = \sqrt{\frac{i\omega}{\alpha_s}}$, where $\alpha_s$ is the thermal diffusivity of steel tube.

The surface temperature rise of a planar plate under the modulated laser heating is:

$$\theta_{s,pl} = \frac{q_s e^{-\frac{\pi}{4}j}}{e_s \sqrt{\omega}} \quad (12)$$

where $e_s$ is thermal effusivity of the plate. **Figure 3b** compares the thermal responses of a cylindrical tube and a plate using the thermal properties of steel in **Table 1**. To quantify the difference between the cylindrical and planar geometries, the ratio $R$ of the $|\theta_s|$ of a cylindrical tube over that of a plate is also displayed in **Figure 3b**, where $R$ is defined as:

$$R = \frac{|\theta_{s,cyl}|}{|\theta_{s,pl}|} \quad (13)$$

The figure shows that the maximum deviation between the two curves is 7% within the entire frequency range we will use in this experiment (0.5 to 2 Hz). Therefore, it is reasonable to use the simplified analysis based on the planar geometry. By neglecting the forced convection effect and temperature gradient in the $x$ direction, and assuming the planar geometry, the frequency-domain heat transfer equation in the fluid domain (**equation 6**) can be further simplified as:

$$\frac{\partial^2 \theta_f}{\partial y^2} = \frac{j\omega}{\alpha_f} \theta_f \quad (14)$$

**Equation 14** is the same as the 1D heat conduction equation in a solid substrate. Therefore, the 1D heat conduction of the three-layered system (i.e., coating, steel containment and fluid, see **Figure 2a**) can be modeled as follows[25, 28]:

$$\theta_s = -\frac{d}{c} q_s \tag{15}$$

where $q_s$ is the AC component of the laser heat flux; $\theta_s = T_s - T_o$ is the temperature oscillation induced by the AC component of the heat flux, with $T_s$ as the transient surface temperature and $T_o$ as the baseline surface temperature due to the heating by the heaters and DC component of laser; $d$ and $c$ are obtained from the following transfer matrix:

$$M = \begin{pmatrix} a & b \\ c & d \end{pmatrix} = M_n M_{n-1} \cdots M_i \cdots M_1 \tag{16}$$

where $M_i$ is the transfer matrix for the $i^{th}$ layer, with $i = 1, 2$ and $3$ referring to the coating, steel tube and fluid, respectively. The transfer matrix $M_i$ is expressed as follows:

$$M_i = \begin{pmatrix} \cosh(D_i\sqrt{j\omega}) & -\frac{\sinh(D_i\sqrt{j\omega})}{e_i\sqrt{j\omega}} \\ -e_i\sqrt{j\omega}\sinh(D_i\sqrt{j\omega}) & \cosh(D_i\sqrt{j\omega}) \end{pmatrix} \tag{17}$$

where $D_i = \frac{l_i}{\sqrt{\alpha_i}}$, $l_i$ and $\alpha_i$ are the thickness and thermal diffusivity of the $i^{th}$ layer, respectively. $e_i$ is the thermal effusivity of the $i^{th}$ layer, i.e., $e_i = \sqrt{(\rho c k)_i}$.

To quantify the forced convection effect on the thermal conductivity measurement, we compare the thermal responses predicted by the stationary fluid model in **equation 15** and that numerically calculated with COMSOL simulation based on **equation 3**. The details of the COMSOL simulation are described in the Supplementary Information Section S2. The modeling parameters are shown in **Table 1**. As shown in **Figure 4a**, at zero bulk velocity ($u_b = 0$ mm s$^{-1}$), the thermal responses of the analytical stationary fluid model (**equation 15**) and the numerical model in COMSOL for the flowing fluid (**equation 3**) agree well with each other. It is evident that the thermal response can be divided into three regions: the fluid dominated region at low frequency where the $|\theta_s|$ $vs$ $\omega^{-1/2}$ curve is linear; the steel dominated region at intermediate frequency where the slope of the $|\theta_s|$ $vs$ $\omega^{-1/2}$ curve becomes smaller because the thermal conductivity of steel is higher than that of fluid; the coating dominated region at high frequency where the slope is large due to the low thermal conductivity of coating. In an earlier study, we have experimentally demonstrated that the thermal conductivity and volumetric heat capacity of different layers can be obtained from the thermal responses at different frequencies [25]. At the low frequency limit when $L_p$ is much larger than the thickness of coating and steel, the multilayer model (**equation 17**) can be simplified as [25, 29]:

$$\theta_s = \frac{q_s e^{-\frac{\pi}{4}j}}{e_f \sqrt{\omega}} + q_s R_{cs} \tag{18}$$

where $e_f$ is the thermal effusivity of fluid, defined as $e_f = \sqrt{k_f \rho_f c_f}$, where $\rho_f$ and $c_f$ are the density and specific heat of the fluid, respectively; $R_{cs}$ is the total thermal resistance by the coating and the steel shell:

$$R_{cs} = \left(1 - \frac{e_s}{e_f}\right)\frac{l_s}{k_s} + \left(\frac{1}{2} - \frac{e_c^2}{e_f^2}\right)\frac{l_c}{k_c} \tag{19}$$

where $e_s$ and $e_c$ are the thermal effusivities of the steel shell and the coating, respectively; $l_s$ and $l_c$ are the thicknesses of the steel shell and the coating, respectively; $k_s$ and $k_c$ are the thermal conductivities of the steel shell and the coating, respectively. Notably, $R_{cs}$ is independent on the frequency in the low frequency range, therefore thermal effusivity of the fluid $e_f$ can be directly obtained by measuring the slope of the $|\theta_s|$ $vs$ $\omega^{-1/2}$ curve according to **equation 19.** At low bulk velocity $u_b = 10$ mm s$^{-1}$ as shown in **Figure 4a**, the thermal response remains the same as that of the stationary fluid. Therefore, the intrinsic thermal effusivity of fluid can still be obtained by measuring the slope at low frequency range ($L_p = 150 – 300$ μm). As $u_b$ increases to above 50 mm s$^{-1}$, the thermal response starts to slightly deviate from that of the stationary fluid. The curves start to plateau at low frequency at $u_b > 100$ mm s$^{-1}$, indicating higher effective thermal effusivity in this regime, or the onset of the forced convection effect. At high flowing velocity, the forced convection effect becomes important, leading to a higher effective thermal conductivity. Therefore, the critical velocity is ~ 100 mm s$^{-1}$ for MPR measurement of intrinsic thermal conductivity of the fluids with the experimental conditions used in this study. The critical velocity predicted by the COMSOL simulation agrees well with that predicted by the dimensional analysis shown in **equation 8**.

**Figure 4b** shows the sensitivity of $|\theta_s|$ on the thermal conductivity of the steel wall $k_s$, Pyromark coating $k_c$ and fluid $k_f$, calculated based on **equation 15** using the parameters in **Table 1**. The sensitivity of $|\theta_s|$ is defined as:

$$S_\theta = \frac{\Delta|\theta_s|/|\theta_s|}{\Delta k_i/k_i}, \tag{20}$$

where $k_i$ is the thermal property of the steel wall, coating, or fluid ($k_s, k_c$, or $k_f$). For example, the sensitivity of 1 to $k_f$ means that 1% change in $k_f$ leads to 1% change in $|\theta_s|$. As discussed above, $|\theta_s|$ is more sensitive to $k_f$ at the low frequency limit (long thermal penetration depth) and to $k_c$ at the high frequency limit (short thermal penetration depth), while in the intermediate frequency range, it is also sensitive to $k_s$. Therefore, in this study, $k_f$ is extracted from the thermal response in the frequency range of 0.5 to 2 Hz ($L_p$ = 150 – 300 μm) where the sensitivity of $|\theta_s|$ to $k_f$ higher than 0.2.

Another important design parameter is the small wall thickness (100 μm) of the MPR section, which not only ensures high sensitivity of $|\theta_s|$ to $k_f$ but also limits the lateral heat spreading effect. **Equation 15** is based on the 1D heat transfer model, which is only valid when there is no significant lateral heat spreading effect, i.e., with small wall thickness. We analyzed the 2D heat transfer effect in the system in Supplementary Information Section S3 and found that when the wall thickness is less than 100 μm, the measurement error of thermal effusivity of the fluid based on the 1D assumption (i.e., **equation 15**) is less than 10%. In addition, to satisfy the 1D assumption, one should also ensure $D_{laser} \gg L_p$ [25]. In our measurement, $L_p$ is ~ 150-300 μm which is much smaller than the $D_{laser}$ (~10 mm). The natural convection effect induced by the temperature gradient is also negligible in the MPR measurement due to the short $L_p$ as discussed in Supplementary Information Section S4. The heat flux is calibrated using a standard material with known thermal effusivity $e$ according to **equation 12**, as shown in Supplementary Information Section

S5. This approach avoids the need to precisely measure the laser spot size which is reported to be a major source of measurement uncertainty[30]. The method of converting the MCT voltage signal to $|\theta_s|$ is also shown in Supplementary Information Section S5. The MPR measurement uncertainty is 5.8% and the analysis can be found in our previous work [25].

## 3. Results and Discussion

### 3.1. Measurement of Stagnant Fluids

**Figure 5a** shows the thermal responses of different fluids at stagnant state (i.e., $u_b$ = 0 mm s$^{-1}$) and room temperature. All the thermal response curves can be delineated into three regions: (i) Pyromark coating dominated region at high frequency with $L_p$ < 30 μm. (ii) tube wall (or substrate) dominated region in the medium frequency range where the slope is smaller due to the high thermal conductivity of the steel. The thermal responses in these two regions are the same for all the samples because the same tube with the same coating was used. (iii) fluid dominated region at low frequency with $L_p$ > ~160 μm, where the slope is determined by the thermal effusivity of the fluid. According to **Equation 18**, the thermal effusivity of the fluid $e_f$ is obtained from the slope of the $|\theta_s|$ $vs.$ $\omega^{-\frac{1}{2}}$ curve in the low frequency regime, and the thermal conductivity of the fluid can be calculated based on the specific heat and density of the fluids reported in the literature [31-33]. **Figure 5b** shows the normalized thermal response by the heat flux ($\frac{\theta_s}{q_s}$) in the frequency range of 0.5-2 Hz, i.e., $L_p$ in the range of 180-320 μm. $e_f$ can be directly obtained from the slopes of the $\frac{\theta_s}{q_s}$ $vs$ $\omega^{-1/2}$ curves. **Figure 5c** compares the measured thermal conductivity of different fluid ($e_{exp}$) to the literature values ($e_{lit}$) while **Figure 5d** compares the measured thermal conductivity of different fluid ($k_{exp}$) to the literature values ($k_{lit}$) [31-33]. The measurement results are within 10% error from the literature values, validating the MPR measurement of stagnant fluids. This error is similar to what we have reported for MPR measurements on bulk materials [25]. The stagnant fluids can be treated as bulk materials if the natural convection effect is negligible. In our measurements, $L_p$ in the fluids ranged from 180 to 320 μm in the low-frequency regime, leading to Rayleigh number ($Ra$) of 1 to 10 for these fluids (the critical $Ra$ for the induction of natural convection is 1100 [34]) and consequently negligible natural convection effect.

## 3.2. Measurement of Flowing Fluids at Room Temperature

According to the analysis presented in Section 2, to measure the thermal conductivity of flowing fluids, the flowing velocity should be lower than the critical velocity to satisfy the condition of negligible forced convection effect. In this section, we measured Xceltherm oil and DI water at room temperature with variable flowing velocities to experimentally identify the critical velocity and compare it to the theoretical analysis. **Figure 6a** shows the thermal responses of Xceltherm oil. The thermal responses for $u_b = 0$ mm s$^{-1}$ and 90 mm s$^{-1}$ coincide, indicating negligible forced convection effect up to 90 mm s$^{-1}$. The thermal response curves shift downwards with increasing flowing velocities, indicating higher effective thermal effusivity and thermal conductivity due to the stronger forced convection effect. The effective thermal effusivity of the fluid ($e_{eff}$) is obtained from the slope of $|\theta_s|$ vs. $\omega^{-\frac{1}{2}}$ in the linear region with $L_p = 150 - 300$ μm (**Equation 18**) and the effective thermal conductivity ($k_{eff}$) is calculated with the literature values of specific heat and density of the fluids. **Figure 6c** shows $k_{eff}$ normalized to the intrinsic thermal conductivity ($k_0$) of Xceltherm oil, $k_{eff}/k_0$, as a function of $u_b$. $k_{eff}/k_0$ is within 5% of the unity when $u_b = 0 - 90$ mm s$^{-1}$. At higher velocity, $k_{eff}/k_0$ increases, reaching 118% at $u_b = 250$ mm s$^{-1}$ and 151% at $u_b = 500$ mm s$^{-1}$. This result indicates that the intrinsic thermal conductivity of Xceltherm oil can be measured at $u_b < 100$ mm s$^{-1}$ for the current MPR configuration, consistent with our theoretical analysis in Section 2.

**Figure 6b** shows the thermal responses of DI water. The slopes of the thermal responses from $u_b = 0$ mm s$^{-1}$ and 90 mm s$^{-1}$ are close, indicating negligible forced convection effect up to 90 mm s$^{-1}$. The thermal response curves shift downwards with increasing velocities, indicating higher effective thermal effusivity and thermal conductivity due to the forced convection, similar to the Xceltherm oil. The curve shows a plateau in low frequency region at $u_b = 550$ mm s$^{-1}$, indicating a significantly higher effective thermal effusivity in this region. **Figure 6c** shows $k_{eff}$ normalized to the intrinsic thermal conductivity ($k_0$) of DI water, $k_{eff}/k_0$, as a function of $u_b$. $k_{eff}/k_0$ is within 5% of the unity when $u_b = 0 - 90$ mm s$^{-1}$. At higher velocity, $k_{eff}/k_0$ increases, reaching 118% at $u_b = 190$ mm s$^{-1}$. The above analysis indicates that the intrinsic thermal conductivity of DI water can also be measured at $u_b < 100$ mm s$^{-1}$, which is close to the prediction from the COMSOL simulation as shown in **Figure 4a** and the dimensional analysis shown in **Figure 3a**. It is also worth noting that the forced convection effect is stronger for DI water than Xceltherm oil at the same flowing velocity. The $k_{eff}/k_0$ reaches 310% at $u_b = 347$ mm s$^{-1}$ and 1130% at $u_b = 550$ mm s$^{-1}$ for DI water (versus 151% at $u_b = 500$ mm for Xceltherm). This is because the viscosity of oil is higher, leading to a lower $Re_D$ number at the same flow velocity, which will be further discussed in the later section.

### 3.3. Measurement of Flowing Fluids at Higher Temperature

**Figure 7a and 7b** show the thermal responses for DI water and Xceltherm 600 oil, respectively, at the same velocity of $u_b = 90$ mm s$^{-1}$ with fluid temperature ($T_0$) ranging from 30°C to 170°C. The upper temperature limit is set so to avoid boiling of water or ignition of oil in air. Like the analysis done in the previous section, $e_{eff}$ of fluid is obtained from the slope at low frequency and then the $k_{eff}$ is calculated. **7d** compares $k_{eff}$ of DI water and the oil to the literature value $k_0$ at different temperature. $k_{eff}$ of DI water is close to $k_0$ from 30°C to 60°C. For the Xceltherm oil, $k_{eff}$ is close to $k_0$ at $T_0 < 100°C$. However, $k_{eff}$ of the oil increases significantly at $T_0 > 100°C$; $k_{eff}/k_0$ is 154.5% and 168.1% at 140°C and 170°C, respectively due to the strong forced convection at higher temperature. The viscosity of the oil decreases with increasing temperature, and thus $Re_D$ increases at higher temperature, leading to stronger forced convection effect. **Figure 7c** shows the thermal responses for the oil at the velocity of $u_b = 50$ mm s$^{-1}$ with $T_o$ from 90°C to 170°C. As shown in **Figure 7d**, by reducing the flow velocity to 50 mm s$^{-1}$, $k_{eff}$ of Xceltherm oil agrees well with $k_0$ at all the temperatures. Evidently, the critical velocity varies with the type of the fluid and the temperature, which will be discussed in the next section.

### 3.4. Analysis of Forced Convection Effect

To systematically study the forced convection effect, we further measured the two fluids at different velocities: (a) Xceltherm oil from 45-170°C with flowing velocity of 50 mm s$^{-1}$, 90 mm s$^{-1}$, 250 mm s$^{-1}$ and 300 mm s$^{-1}$; and (b) DI water at 30 °C with flowing velocity of 90 mm s$^{-1}$, 190 mm s$^{-1}$, 350 mm s$^{-1}$ and 550 mm s$^{-1}$. **Figure 8a** shows the normalized thermal conductivity ($k_{eff}/k_0$) from all the measurements of DI water and Xceltherm 600 as a function of $Re_D$. It can be seen that $k_{eff}/k_0$ increases with increasing $Re_D$. Although the range of velocity is the same for DI water and Xceltherm oil, $Re_D$ for DI water is much higher due to the lower viscosity. At $u_b$ > 190 mm s$^{-1}$, $Re_D$ > 10,000, the flow of the water becomes turbulent, leading to strong forced convection effect. Therefore, $k_{eff}$ also increases significantly. The viscosity of the oil decreases with increasing temperature, and thus $Re_D$ is higher at higher temperature. As a result, $k_{eff}$ directly obtained from the measurement is increasingly overestimated compared to $k_0$ as the temperature increases, as shown in **Figure 7d**. However, compared to DI water, $Re_D$ for Xceltherm is less than 10,000 and the flow is still laminar within the entire flow velocity range.

Here we plot $k_{eff}/k_0$ as a function of $Re_D^{1/2}Pr^{1/3}$ in **Figure 8b**. Notably, $k_{eff}/k_0$ of both DI water and Xceltherm 600 follows the same trend as $Re_D^{1/2}Pr^{1/3}$ increases, except the outliner for DI water at high $Re_D^{1/2}Pr^{1/3}$ value (>200) because the water flow is turbulent, which does not follow the $Nu_D$ correlation for a laminar flow. The forced convection effect is negligible for both liquids at $Re_D^{1/2}Pr^{1/3}$ < 100 and becomes increasingly prominent as $Re_D^{1/2}Pr^{1/3}$ increases. This agrees well with the theoretical analysis as shown in **equation 10**. Therefore, we conclude that it is feasible to use the MPR technique developed here to measure the intrinsic thermal conductivity of flowing fluids under the condition of both low $u_b$ and low $Re_D^{1/2}Pr^{1/3}$ (or more generally, **equation 7** and **equation 11** are both satisfied). On the other hand, this work also suggests that the MPR could be used to probe the convection effect of flowing fluids when $u_b$ or $Re_D^{1/2}Pr^{1/3}$ is larger.

## 4. Conclusion

In this work, we develop an MPR setup to measure the thermal conductivity of flowing fluid in a tube by controlling the modulation frequency of heating laser beam and the corresponding thermal penetration depth. The MPR setup is first validated by the thermal conductivity measurement of stationary fluids, including DI-water, Dowtherm A oil, Xceltherm 600 oil and ethanol at room temperature. The measurement error of stationary fluid is within 10%. Subsequently, the MPR setup is extended for the measurement of flowing fluid in the velocity range of 0 – 550 mm s$^{-1}$ and from room temperature to 170°C. The intrinsic thermal conductivity of fluid can be obtained at low velocity where the forced convection effect is minimized. The forced convection effect becomes prominent at high flow velocity and low frequency, leading to the overestimation of thermal conductivity. The critical velocity is determined by the dimensionless number $Re_D^{1/2} Pr^{1/3}$ modeled based on the laminar boundary layer. When $Re_D^{1/2} Pr^{1/3} < 100$, the intrinsic thermal conductivity of fluid can be obtained. The MPR technique reported here provides a convenient tool of *in-situ* thermal conductivity measurement of flowing fluid and shows the potential to characterize the local forced convection heat transfer close to the wall.

**Supplementary Information**

Photo of the MPR system; COMSOL simulation of flowing fluid; Analysis of 2D heat transfer effect in the MPR measurement; Analysis of natural convection effect in the MPR measurement; Method of heat flux and temperature calibration.

**Acknowledgements**

This material is based upon work supported by the U.S. Department of Energy's Office of Energy Efficiency and Renewable Energy (EERE) under Solar Energy Technologies Office (SETO) Agreement Number DE-EE0008379. The views expressed herein do not necessarily represent the views of the U.S. Department of Energy or the United States Government.

**Author contribution statement**

**Figure and Table Captions**

Figure 1. MPR system integrated with a flowing fluid loop. (a) Overview of the system. Schematic (b) and photograph (c) of the MPR measurement section.

Figure 2. Schematic of MPR measurement on flowing fluid in a tube. (a) Schematic of the MPR measurement section. (b) Schematics of momentum boundary, thermal boundary and thermal penetration depth in the MPR measurement in a 2D model.

Figure 3. Simplification of heat transfer modeling in the fluid domain. (a) Configuration of MPR measurement on flowing fluid for minimum forced convection effect. (b) Comparison of thermal responses between the cylindrical and planar geometries. The left *y*-axis is the amplitude of surface temperature oscillation and the right *y*-axis (*R*) is the ratio of the $|\theta_s|$ between the cylindrical and planar geometries.

Figure 4. Modeling of thermal response of flowing fluid and measurement error analysis of the thermal conductivity of fluid due to the forced convection. (a) Effect of flowing velocity $u_b$ on the thermal response. The solid line from the analytical model is based on **equation 16** and the symbols are results from finite element modeling. (b) Sensitivity on the thermal conductivities of coating, steel tube and fluid at $u_b = 0$ mm s$^{-1}$.

Figure 5. MPR measurement on stationary fluids at room temperature. (a) Thermal responses of different stagnant fluids filled in the MPR measurement section (tube wall thickness: 100 μm; Pyromark coating thickness: ~ 30 μm). $L_P$ is estimated based on the properties of DI water. (b) Thermal responses of different stagnant fluids at the low frequency range of 0.5-2 Hz. (c) Comparison of the thermal effusivity measured using MPR and that reported in the literature. (d) Comparison of the thermal conductivity measured using MPR and that reported in the literature.

Figure 6. MPR measurement on flowing fluids at room temperature. (a) Thermal response of Xceltherm 600 oil in a tube flowing at different velocities. (b) Thermal response of DI water in a tube flowing at different velocities. (c) Normalized thermal conductivity as a function of flow velocity for DI water and Xceltherm 600 oil. The thermal effusivity measured from the slopes of the thermal response curves are used to calculate the thermal conductivity based on the density and specific heats in the literature.

Figure 7. MPR measurement of flowing fluids at elevated temperature. (a) Thermal responses of DI water at $u_b$ = 90 mm s$^{-1}$ at 30-60°C. (b) Thermal response of Xceltherm 600 oil at $u_b$ = 90 mm s$^{-1}$ at 45-170°C. (c) Thermal response of Xceltherm 600 oil at $u_b$ = 50 mm s$^{-1}$ at 90-170°C. (d) Effective thermal conductivities of DI water and Xceltherm 600 oil as a function of temperature measured at $u_b$ = 90 mm s$^{-1}$ and $u_b$ = 50 mm s$^{-1}$.

Figure 8. Analysis of forced convection effect in flowing fluid. (a) Normalized thermal conductivity as a function of $Re_D$ number for DI-water and Xceltherm 600 oil. (b) Normalized thermal conductivities of DI water and Xceltherm 600 oil as a function of $Re_D^{1/2} Pr^{1/3}$ at different temperatures.

Table 1. Simulation parameters for stationary fluid in a tube

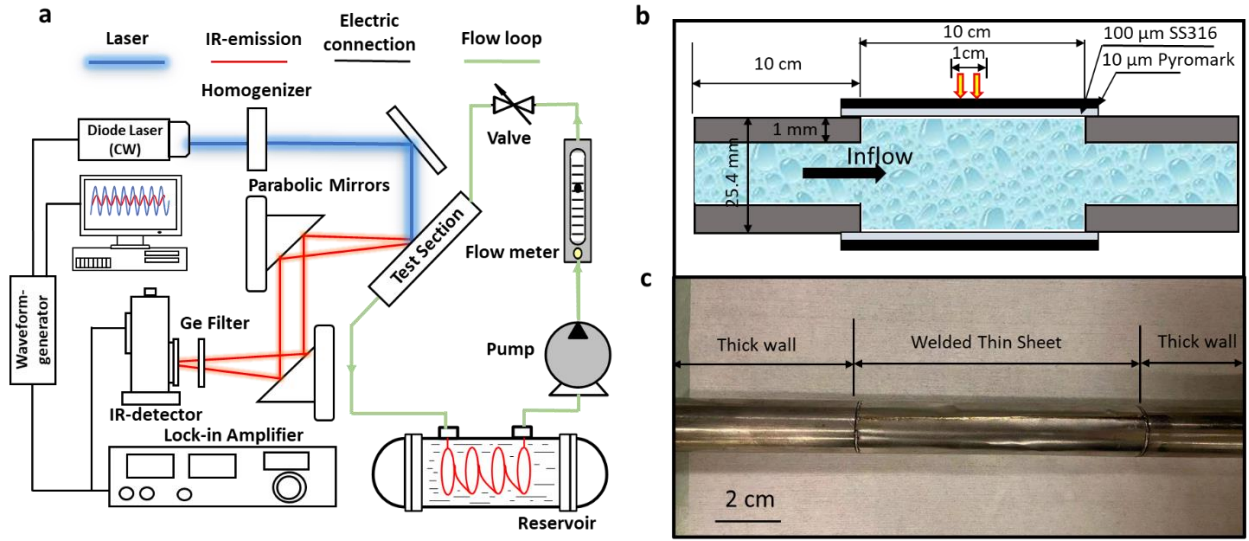

**Figure 1.** MPR system integrated with a flowing fluid loop. (a) Overview of the system. Schematic (b) and photograph (c) of the MPR measurement section of the pipe for the fluid flow.

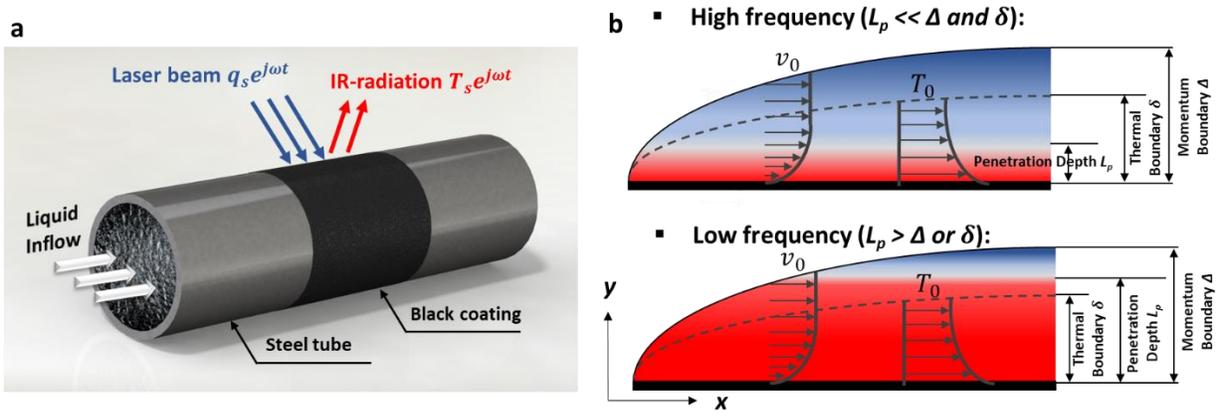

**Figure 2.** Schematic of MPR measurement on flowing fluid in a tube. (a) Schematic of the MPR measurement section. (b) Schematics of momentum boundary, thermal boundary and thermal penetration depth in the MPR measurement in a 2D model.

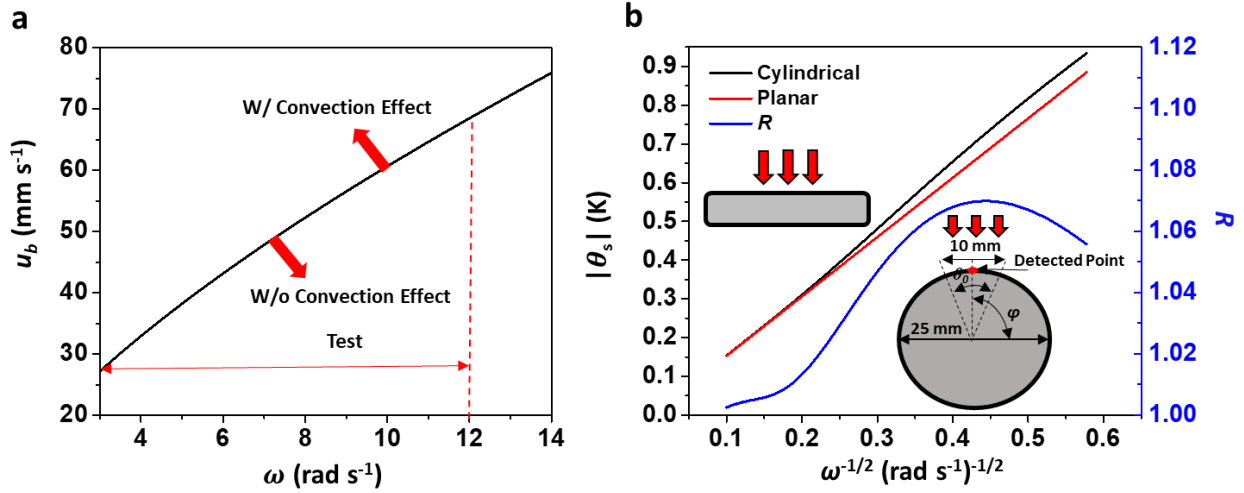

**Figure 3.** Simplification of heat transfer modeling in the fluid domain. (a) Configuration of MPR measurement on flowing fluid for minimum forced convection effect. (b) Comparison of thermal responses between the cylindrical and planar geometries. The left *y*-axis is the amplitude of surface temperature oscillation and the right *y*-axis (*R*) is the ratio of the |$\theta_s$| between the cylindrical and planar geometries.

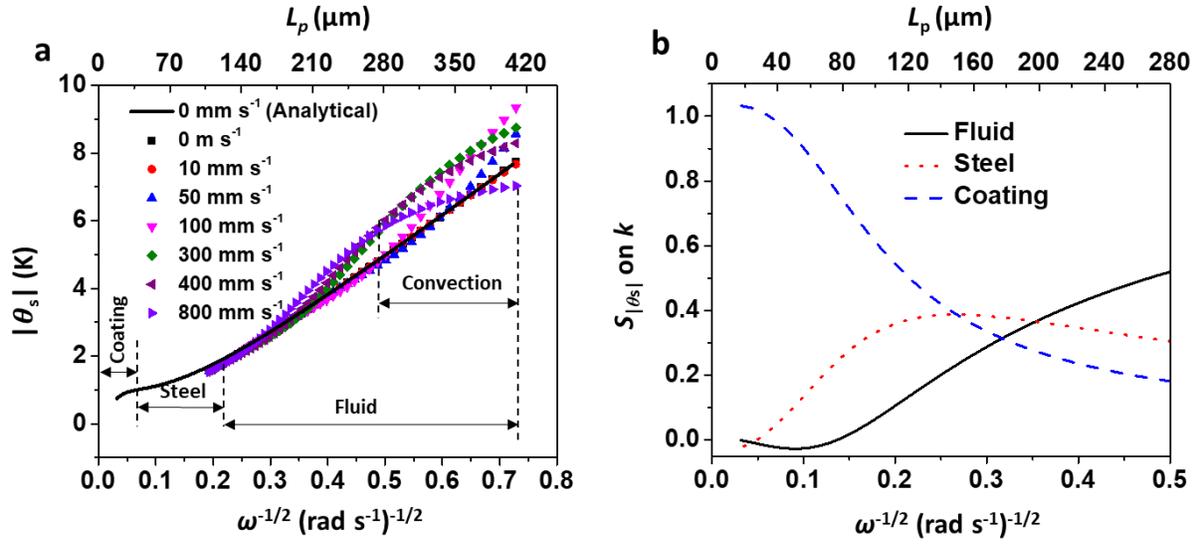

**Figure 4.** Numeric modeling of thermal response of flowing fluid and measurement error analysis of the thermal conductivity of fluid due to the forced convection. (a) Effect of flowing velocity $u_b$ on the thermal response. The solid line from the analytical model is based on **equation 15** and the symbols are results from finite element modeling using COMSOL. (b) Sensitivity on the thermal conductivities of coating, steel tube and fluid at $u_b = 0$ mm s$^{-1}$.

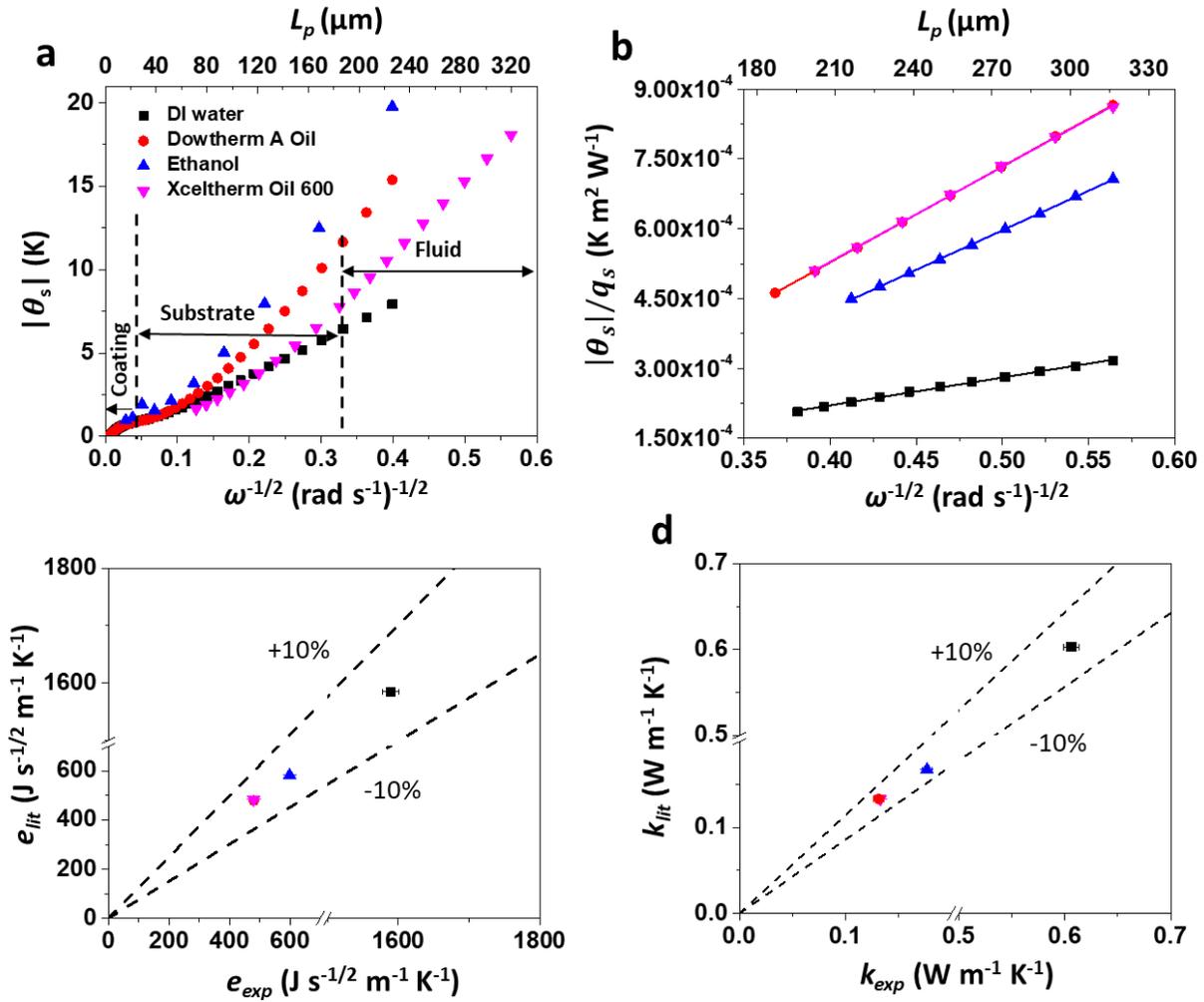

**Figure 5.** MPR measurement on stationary fluids at room temperature. (a) Thermal responses of different stagnant fluids filled in the MPR measurement section (tube wall thickness: 100 μm; Pyromark coating thickness: ~ 30 μm). $L_P$ is estimated based on the properties of DI water. (b) Thermal responses of different stagnant fluids at the low frequency range of 0.5-2 Hz. (c) Comparison of the thermal effusivity measured using MPR and the literature values. (d) Comparison of the thermal conductivity measured using MPR and the literature values. Note the data of Dowtherm A oil and Xcetherm oil 600 are overlapped in (b), (c), and (d).

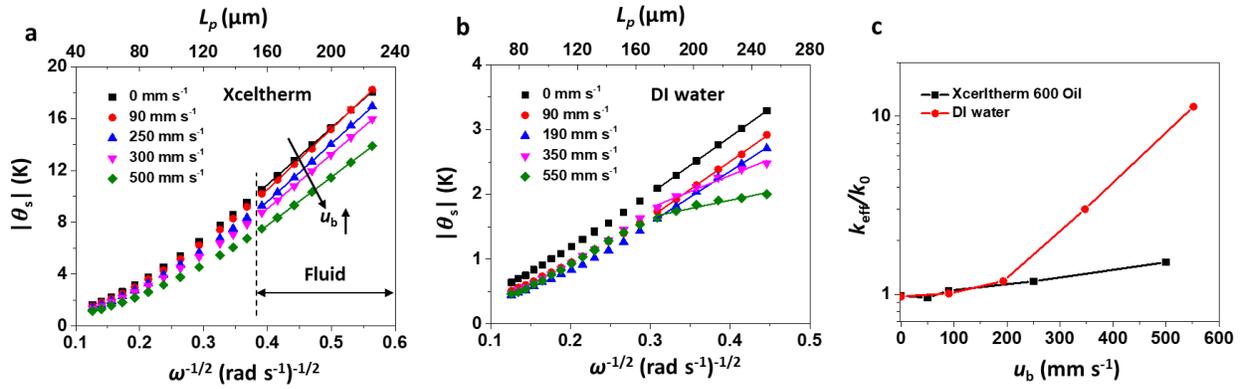

**Figure 6.** MPR measurement on flowing fluids at room temperature. (a) Thermal response of Xceltherm 600 oil in a tube flowing at different velocities. (b) Thermal response of DI water in a tube flowing at different velocities. (c) Normalized thermal conductivity as a function of flow velocity for DI water and Xceltherm 600 oil. The thermal effusivity measured from the slopes of the thermal response curves are used to calculate the thermal conductivity based on the density and specific heats in the literature [31-33].

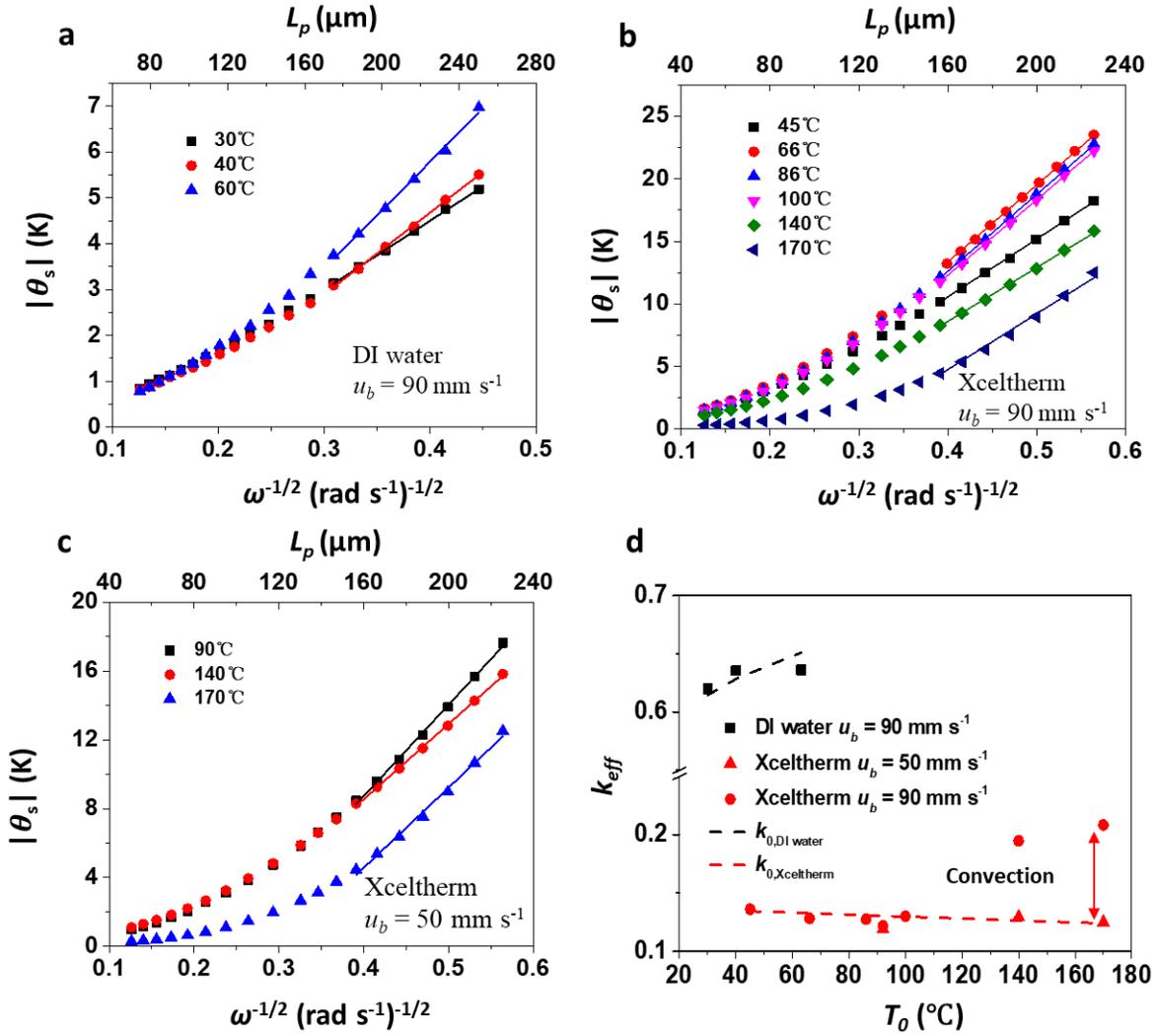

**Figure 7.** MPR measurement of flowing fluids at elevated temperature. (a) Thermal responses of DI water at $u_b$ = 90 mm s$^{-1}$ at 30-60°C. (b) Thermal response of Xceltherm 600 oil at $u_b$ = 90 mm s$^{-1}$ at 45-170°C. (c) Thermal response of Xceltherm 600 oil at $u_b$ = 50 mm s$^{-1}$ at 90-170°C. (d) Effective thermal conductivities of DI water and Xceltherm 600 oil as a function of temperature measured at $u_b$ = 90 mm s$^{-1}$ and $u_b$ = 50 mm s$^{-1}$.

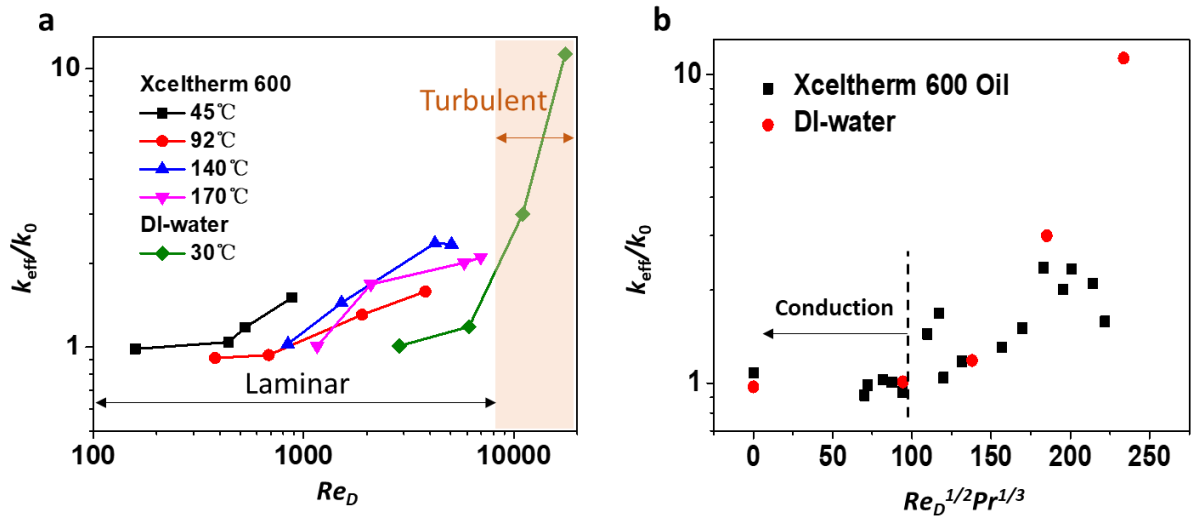

**Figure 8.** Analysis of forced convection effect in flowing fluid. (a) Normalized thermal conductivity as a function of $Re_D$ number for DI-water and Xceltherm 600 oil. (b) Normalized thermal conductivities of DI water and Xceltherm 600 oil as a function of $Re_D^{1/2} Pr^{1/3}$ at different temperatures.

**Table 1.** Simulation parameters for stationary fluid within a tube

| Parameter | Value | Unit |
|---|---|---|
| Tube Diameter $D$ | 25 | mm |
| Thickness of Tube Shell $l_s$ | 100 | μm |
| Thickness of Coating $l_c$ | 30 | μm |
| Thermal Conductivity of Steel Tube $k_s$ | 20 | W m$^{-1}$ K$^{-1}$ |
| Thermal Conductivity of Coating $k_c$ | 0.5 [25] | W m$^{-1}$ K$^{-1}$ |
| Thermal Conductivity of Fluid $k_f$ | 0.6 | W m$^{-1}$ K$^{-1}$ |
| Thermal Diffusivity of Tube $α_s$ | 5 | mm$^2$ s$^{-1}$ |
| Thermal Diffusivity of Coating $α_c$ | 0.4 | mm$^2$ s$^{-1}$ |
| Thermal Diffusivity of Fluid $α_f$ | 0.16 | mm$^2$ s$^{-1}$ |
| Heat Flux $q_s$ | 13700 | W m$^{-2}$ |
| Heating Laser Diameter $D_{\text{Laser}}$ | 10 | mm |
| Detection Spot Diameter $D_{\text{IR}}$ | 1 | mm |